\documentclass[12pt,preprint]{aastex}

\slugcomment{Accepted for publication: Astronomical Journal}

\shorttitle{NGC 6907/8}
\shortauthors{Madore et al.}

\begin{document}


\title{The Curious Case of NGC~6908\footnote{Based on data obtained
at the Magellan and duPont telescopes which are operated by the
Carnegie Institution of Washington}}


\author{Barry F. Madore}
\affil{Observatories of the  Carnegie Institution of Washington \\ 813 Santa 
Barbara St., Pasadena, CA ~~91101}
\author{Armando Gil de Paz}
\affil{Dept. de Astrofisica, Universidad Compluetense, Madrid E-28040, Spain}
\author{Olga Pevunova}
\affil{NASA/IPAC EXtragalactic Database, California Institute of Technology \\ IPAC MS 100-22, Pasadena, CA ~~91125}
\author{Ian Thompson}
\affil{Observatories of the  Carnegie Institution of Washington \\ 813 Santa 
Barbara St., Pasadena, CA ~~91101}
\email{barry@ociw.edu, agpaz@astrax.fis.ucm.es, olga@ipac.caltech.edu,ian@ociw.edu}



\begin{abstract}
The object NGC~6908 was once thought to be simply a surface-brightness
enhancement in the eastern spiral arm of the nearby spiral galaxy
NGC~6907.  Based on an examination of near-infrared imaging, the
object is shown in fact to be a lenticular S0(6/7) galaxy hidden in
the optical glare of the disk and spiral structure of the larger
galaxy.  New radial velocities of NGC~6908 (3,060 $\pm$ 16 (emission);
3,113 $\pm$ 73 km/s (absorption)) have been obtained at the Baade 6.5m
and the duPont 2.5m telescopes at Las Campanas, Chile placing NGC~6908
at the same expansion-velocity distance as NGC~6907 (3,190 $\pm$ 5
km/s), eliminating the possibility of a purely chance line-of-sight
coincidence. The once-enigmatic asymmetries in the disk and outer
spiral structure of NGC~6907 are now explained as being due to an
advanced merger event. Newly discovered tails and debris in the outer
reaches of this galaxy further support the merger scenario for this
system. This pair of galaxies is a rather striking example of two
objects discovered over 100 years ago, whose true nature was lost
until modern detectors operating at infrared wavelengths gave us a
new (high-contrast) look. Other examples of embedded merger
remnants may also reveal themselves in the growing samples of
near-infrared imaging of nearby galaxies; and a pilot study does
reveal several other promising candidates for follow-up observations.
 ~

~

~

\end{abstract}

\section{INTRODUCTION}

In his General Catalogue of nebulae, Sir John Herschel (1864)
discussed two non-stellar, extended objects which were later added to
the compilation of Dreyer (1888) and given their long-lasting
designations, NGC~6907 and NGC~6908. Translated from the shorthand
notation as published \footnote{ In the cryptic empirical notation of
the time, NGC~6907 was described as being {\tt (cF, cL, vlE, vglbM, r,
3 st p)}, while the entry for NGC~6908 was {\tt (eF, vS, lE, h
2076 p)}.}  NGC~6907 was described as {\it ``considerably faint,
considerably large, very little extended, very gradually a little
brighter towards the middle, mottled, 3 stars preceding''}, and the
eastern object, NGC~6908 was said to be {\it ``extremely faint, very
small, little extended, Herschel 2076 (aka NGC~6907) preceding''.}

On the sky, the two objects discussed by Herschel are found only
$\sim$40~arcsec apart; a close pair by any standards, but otherwise
unremarkable.  With time these objects were observed again, not
directly at the eyepiece of the telescope as was done by Herschel, but
only after having been recorded on photographic plates and then
visually re-inspected by eye.  What was revealed at that point in time
was a single much more expansive parent object, the galaxy NGC~6907,
having a ``massive'' arm, asymmetrically placed to the east, and
`certainly' to be identified with NGC~6908. From this point onward
this nebulous object to the east lost its independent status and was
absorbed into the galactic sub-structure of NGC~6907.  Subsequent
observers and classifiers agreed (e.g., Nilson 1974) that the dominant
spiral galaxy NGC~6907 had an ``asymmetric disk'' and with ``two
massive, asymmetric arms'', the more prominent one being seen to the
east. After an abbreviated search for a cause of this imbalance Nilson
made the point of saying in his notes that there was ``no disturbing
object visible.''  This of course suggests that the asymmetry was
generated internally.

In NASA/IPAC Extragalactic Database (NED), NGC~6907 is designated as
SB(s)bc (from the RC3, de Vaucouleurs, et al. 1991) and has a
published optical radial velocity of 3,186 $\pm$8~~km/s (da Costa et
al. 1991). A search of NED, over a circular area one degree ($\sim$
700~kpc) in radius, surrounding the galaxy reveals no objects with
cataloged radial velocities near to NGC~6907 itself. Locally NGC~6907
is quite isolated \footnote{The three closest ``radial velocity
confirmed'' companions within a projected distance of 1~Mpc from
NGC~6907, are IC~5005, ESO~527-G019 and IC~4999. They are
respectively, 61 arcmin (760 kpc), 72 arcmin (890 kpc) and 74 arcmin
(920 kpc), away from NGC~6907.}, although it is associated more widely
with IC~4995 and IC~5005 in the group catalog of Maia, da Costa \&
Latham (1989).\footnote{We note in passing, that a search over an even
larger radial distance reveals a rather peculiarly cold structure: a
sheet of galaxies ten degrees ($\sim$7~Mpc) in diameter all clustered
at $\sim$3,100~km/s, having an rms velocity dispersion about the mean
of only $\pm$52~km/s. In addition to NGC~6907 itself, the galaxies
making up this sheet are IC~5005 [3,112~km/s], ESO~527-G019
[3,117~km/s], IC~4999 [3,156~km/s], AM~2029-235 NED01 [3,065~km/s],
AM~2029-235 NED02 [3,061~km/s], ESO~462-G025 [3,057~km/s],
ESO~462-G016 [3,054~km/s], ESO~462-G028 [3,017~km/s], ESO~596-G030
[3,110~km/s]. }

Until recently NGC~6908 was classified in NED as being a PofG (= Part
of a Galaxy) consistent with the above comments that it was a
high-surface-brightness part of the asymmetric eastern arm of
NGC~6907.  A picture of NGC~6907 is shown in Figure~1 (upper right panel)
taken in the B band.  Outside of the general catalogs there are no
references in the modern literature to NGC~6908, except for its
inclusion (without comment) in a search for companions to barred
galaxies undertaken by Garcia-Barreto, Carrillo \& Vera-Villamizar
(2003) using NED (after the reclassification), and the enigmatic
identification of NGC~6908 with NGC~6907 in RC1 (de Vaucouleurs \& de
Vaucouleurs 1961). For all intents and purposes then, NGC~6908 had
disappeared from the astronomical literature for the better part of
the last century.

\section{ NGC~6908 Redux}

The primary objective of the present study is to re-affirm the
individuality of NGC~6908 and to prove that it is indeed a galaxy in
its own right, albeit one heavily implicated with the disk and spiral
structure of NGC~6907. The first indication of its true identity came
from multi-color imaging.\footnote{CCD imaging data of NGC~6907/8 used
in this work are publicly available through the Isaac Newton Group
(ING; UBRI bands) and Anglo-Australian Telescope (AAT; H$\alpha$ and
H$\beta$) archives, the OSU Bright Spiral Galaxy Survey (deep H band;
Eskridge et al. 2002), and the 2MASS Large Galaxy Atlas (JHKs bands;
Jarrett et al. 2003).}

Panels in Figure~1 progressively reveal NGC~6908 for what it is, and
laterally show why it was so easy to mistake this object for an
enhancement of the eastern arm of NGC~6907. The upper left panel of
Figure~1 shows what observers using B-band photographic plates would
have been confronted with. As one moves progressively to longer and
longer wavelength images the color contrast between NGC~6908 (an
intrinsically red object) and the surrounding (intrinsically blue)
spiral structure increases. By the time the I-band image is seen the
identity of NGC~6908 is already clear. Indeed, the immediate visual
impression from the near-infrared image is that NGC~6908 is a very
early-type galaxy, an E or S0 of rather large apparent flattening.
Isophotometry of images at all available wavelengths confirm and
quantify this impression: outside of the central 4 arcseconds (where
seeing effects are still important) the axial ratio a/b settles down
to a value in the range 0.3-0.4 as measured in all seven bandpasses
from U to K. This gives an integer ellipticity of 6-7.

\vfill\eject
\section{NGC~6907/8: A Newly-Discovered Interacting Pair}

Having established the general nature of NGC~6908 the question
naturally arises: Is this object a chance projection of a background
galaxy along the line of sight, or is it truly involved in an on-going
interaction with the large spiral, NGC~6907? The asymmetry in the arm
structure in NGC~6907 (NGC~6908 to the contrary) is still very
suggestive of an interaction event, be it past or present.

The surface brightness profile of NGC~6908 is very close to a pure
exponential at all wavelengths (see Figure~2). This suggests that
NGC~6908 is not a luminous elliptical (background or otherwise) but
more likely a lenticular S0 system or even possibly a dwarf elliptical
galaxy. However, the high extrapolated central surface brightness
($\mu_{B,0}$=19.6~mag/arcsec$^{2}$; see Binggeli \& Cameron 1991) and
the lack of a significant nuclear excess, common to the majority of
dwarf ellipticals of this luminosity (Ferguson \& Binggeli 1994),
favor the classification of NGC~6908 as a lenticular S0(6/7) galaxy.

In order to finally exclude the possibility of NGC~6908 being a
projection of a foreground or background object along the line of
sight of NGC~6907, on the night of 01 August 2002 we determined its
recession velocity. On that date an optical spectrum was obtained at
the 6.5m Baade telescope at Las Campanas (Chile) using the B\&C camera
with a 1200-line grating and a 0.9-arcsec-wide slit. Our 10~min
exposure quickly revealed a hydrogen absorption-line spectrum
including forbidden [OII] emission at 3727$\AA$; and a subsequent
reduction with respect to a K-star template gave a heliocentric radial
velocity of 3,113 $\pm$ 73~km/s.  Since the parent galaxy NGC~6907 has
a published radial velocity of 3,161~km/s (RC3) this was immediate and
conclusive proof that the two galaxies are coincident in space. This
conclusion was also confirmed by the presence of H$\alpha$ and
possibly also H$\beta$ emission arising from the nucleus of NGC~6908
as seen in archival narrow-band images (tuned to the redshift
NGC~6907) obtained at the AAT (Figure~3).

Follow-up observations of both NGC~6907 and NGC~6908 were later made
at the duPont 2.5m telescope, also at Las Campanas, but this time
using the long-slit capabilities of the Wide-Field camera and
spectrograph so as to simultaneous measure the absolute and
differential velocities of the two galaxies. Those spectra are shown
in Figure~4. The radial velocity derived from the emission lines in
the spectrum of NGC~6908 is 3,060 $\pm$ 16~km/s, while for the nucleus
of NGC~6907 we obtain 3,190 $\pm$ 5~km/s. This yields a velocity
difference between the two systems of 130 $\pm$ 17~km/s. For
H$_0$=70\,km/s/Mpc the absolute magnitude of NGC~6908 is
M$_B$=$-$17.4~mag. The high [NII]$\lambda$6583\,\AA/H$\alpha$
($\sim$0.9) and [SII]$\lambda\lambda$6717,6731\,\AA\AA/H$\alpha$
($\sim$1.7) line ratios and relatively narrow emission lines (FWHM $<$
250~km/s) found in the spectrum of NGC~6908 suggest the existence of a
Seyfert 2 or more probably (considering the high intensity of the
[SII] doublet) a LINER nucleus in this galaxy (e.g. Ho, Filippenko, \&
Sargent 1997). Oxygen lines traditionally used in the classification
of active nuclei ([OII]$\lambda\lambda$3726,3729\,\AA\AA,
[OIII]$\lambda$5007\,\AA, [OI]$\lambda$6300\,\AA) were not detected by
our spectroscopic observations. Although these line ratios are
compatible with emission arising from a supernova remnant (SNR), the
H$\alpha$ luminosities of SNR (especially those with high
[SII]/H$\alpha$ line ratios; see Blair \& Long 1997) are well below
the detection limits of both our imaging and spectroscopic
observations. Indeed, from the H$\alpha$ equivalent width measured in
the spectrum of NGC~6908 (Figure~4), $\sim$4\,\AA, and the $R$-band
surface brightness in the nuclear regions of this galaxy
($\sim$18.5\,mag\,arcsec$^{-2}$; Figure~2) we estimate the H$\alpha$
luminosity of NGC~6908 (within the slit) to be roughly
$\sim$10$^{38.8}$\,erg\,s$^{-1}$, which is well within the range of
luminosities found in nearby LINERs (Terashima, Ho, \& Ptak 2000).

The sum of evidence, both kinematic and morphological, suggests that
the asymmetry in the disk and in the arm structure of NGC~6907 is due
to an on-going and rather highly advanced interaction with the
low-luminosity S0(6/7) galaxy NGC~6908 at a (projected) velocity
difference of $\sim$130~km/s. And, we suggest, that the rather rare
form ({\it i.e.,} the large ellipticity) and moderate nuclear activity
of NGC~6908 are also likely due to the interaction. Figure~5 shows a
RGB composite map produced by combining the B (blue), R (green),
H-band (red) images of the NGC~6907/8 system. In addition to the dust
lanes seen along the leading edges of the spiral arms (common in
strongly barred Sb and Sbc galaxies like NGC~6907) this figure shows a
dust lane crossing the western spiral arm of NGC~6907 almost
perpendicular to it and extending well into the inter-arm region to the
south of the galaxy nucleus (see also Sandage \& Bedke 1994). We
speculate that this feature also might well also be a by-product of
the interaction between NGC~6907 and NGC~6908.

Beyond the high-surface-brightness inner disk of NGC~6907 there is
further evidence that a merger is underway. Figure 6 shows a 10
$\times$ 10~arcmin image of the NGC~6907/8 pair cut from the Digital
Sky Survey and heavily stretched so as to emphasize very
low-surface-brightness features. As can be readly seen the evidence
for an earlier phase in the interaction can be seen in the
asymmetrically placed tail of debris found to be extending out from
the northern part of the main body of the galaxy to the west and then
down to south-west. Even lower surface brightness regions can also be
seen to the east folding over the galaxy clock-wise to the north. Had
the galaxy NGC~6907 been inspected first at this low a
surface-brightness level it is likely that a merger event would have
been independently suggested. Figure 7 is a confirming image taken
with a CCD camera mounted on the 2.5m duPont telescope on Las
Campanas, Chile. The tidal feature, although very faint and close in
surface brightness to the night sky, is visible both in the blue
(photographic) and in the red (R-band CCD) images, confirming its
reality and general structure.

\vfill\eject
\section{Conclusions}

Once thought to be a relative brightening in the ``massive'' eastern
spiral arm in the asymmetric spiral galaxy NGC~6907, NGC~6908 is
revealed by near-infrared imaging to be, in fact, a low-luminosity
(M$_B$=$-$17.4~mag) S0(6/7) galaxy, embedded in the disk of and
strongly interacting with its larger spiral galaxy host,
NGC~6907. This on-going merger of two systems is almost certainly
responsible for the observed asymmetries in the disk and the spiral
structure of NGC~6907, as well as the strong, asymmetrical (tidal
debris) tails seen at low light levels beyond the main body of the
spiral. The present-day interaction may also plausibly account for the
unusually large ellipticity and mild nuclear activity of NGC~6908
itself.

If this pair of galaxies was mistakenly overlooked, are there more
examples of embedded companions in the disks of nearby spirals? It
seems reasonable to suppose that there are some less spectacular
examples, and some more highly advanced (more deeply embedded)
on-going mergers still awaiting revelation, even in relatively nearby
galaxies already well studied at blueward wavelengths.  As more
near-infrared imaging of nearby galaxies comes available it will be of
interest to see whether this serendipitous discovery of an embedded
merger remnant generalizes to a larger number of incidents or whether
this is such a short-lived part of a rare event that it alone is our
only local example. 

Anticipating the results of a larger and more comprehensive survey
(Madore 2007, in preparation) we can say that the answer to the above
question is yes. The larger survey is based upon the simultaneous
visual inspection and intercomparison of both the near-infrared and
the optical images of the entire NGC galaxy sample. A few early
examples, IC~5135, NGC~0238, NGC~5430 and NGC~5534, are given in
Figure~8. These were found by inspecting the 2MASS near-infrared JHKs
images as made available by NED, many of which were originally
published by Jarrett, et al. (2003) and comparing them with on-line
versions of their optical images from the POSS. New wavelengths reveal
new phenomena.

The last word goes to de Vaucouleurs \& de Vaucouleurs (RC1: 1964) who
comment that NGC~6907 is {\it ``slightly asymmetric, and similar to
NGC~1097''.} The irony, of course, is that the asymmetry in NGC~1097
was, even then, clearly due to its interaction with its elliptical
companion NGC~1097A; the latter being a radial-velocity companion
falling within 100~km/s of its parent galaxy and being clearly visible
on any wide-field image of the system. And so now the analogy is
confirmed; at low surface-brightness levels NGC~1097 shows many of the
same signs of tidal debris coming from the earlier interaction phase
as in NGC~6907. Had the companion, NGC~1097A, been along the same line
of sight as one of the main spiral arms of NGC~1097 the correspondence
between NGC~1097 and NGC~6907 would have been exact.

\medskip

\centerline{\it Acknowledgments} We thank Harold Corwin for his
comments on early descriptions of these galaxies. This research made
use of the NASA/IPAC Extragalactic Database (NED) which is operated by
the Jet Propulsion Laboratory, California Institute of Technology,
under contract with the National Aeronautics and Space Administration.

\vfill\eject
\noindent
\centerline{\bf References \rm}
\vskip 0.1cm
\vskip 0.1cm

\par\noindent
Binggeli, B., \& Cameron, L.M. 1991, A\&A, 252, 27

\par\noindent
Blair, W.P., \& Long, K.S. 1997, ApJS, 108, 261

\par\noindent
da Costa, N.L., Davis, M., Meiksin, A., Sargent, W.L.W. \& Tonry, J.L.
1991, ApJS, 75, 935

\par\noindent
de Vaucouleurs, G. \&  de Vaucouleurs, A. 1964
Reference Catalogue of Bright Galaxies,  Austin: University of Texas (RC1)

\par\noindent 
de Vaucouleurs, G. \& de Vaucouleurs, A., Corwin,
Jr. H.G., Buta, R.J., Paturel, G. \& Fouque, P. 1991 Third Reference
Catalogue of Bright Galaxies, New York: Springer-Verlag (RC3)

\par\noindent
Dreyer, J.L.E., 1888 Mem RAS, 49, 1

\par\noindent
Eskridge, P.B., et al. 2002, ApJS, 143, 73

\par\noindent
Ferguson, H.C., \& Binggeli, B. 1994, A\&ARv, 6, 67

\par\noindent
Garcia-Barreto, J.A., Carrillo, R. \& Vera-Villamizar, N. 2003 AJ, 126, 1707

\par\noindent
Herschel, J., 1864 Philosophical Transactions

\par\noindent
Ho, L.C., Filippenko, A.V., \& Sargent, W.L.W. 1997, ApJS, 112, 315

\par\noindent
Jarrett, T. H., Chester, T., Cutri, R., Schneider, S.E., \& Huchra,
J.P. 2003 AJ, 125, 525

\par\noindent
Maia, M.A.G., da Costa, L.N. \& Latham, D.W. 1989 ApJS, 69, 809

\par\noindent
Mathewson, D.S. \& Ford, V.L. 1996, ApJS, 107, 97

\par\noindent 
Nilson, P. 1974 Catalogue of Selected Non-UGC Galaxies,
Uppsala Obs. Report 5

\par\noindent 
Sandage, A.R., \& Bedke, J. 1994 The Carnegie Atlas of Galaxies. 
Volume I (Carnegie Institution of Washington)

\par\noindent 
Terashima, Y., Ho, L.C., \& Ptak, A.F. 2000, ApJ, 539, 161


\vfill\eject
\vskip 0.75cm

 
\begin{figure}
\epsscale{0.5}
\plotone{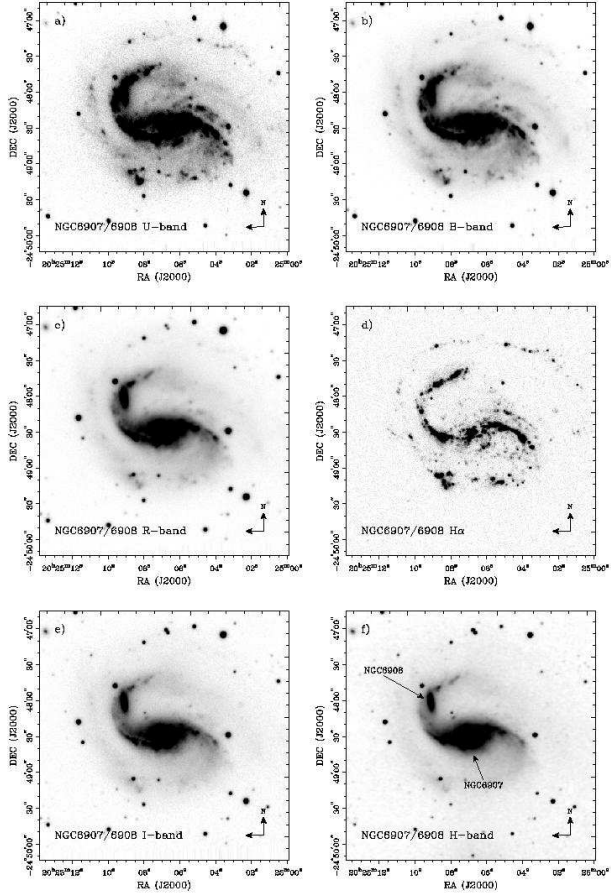}
\vfill
\caption{A multi-wavelength comparison of
images of NGC~6907 and NGC~6908 from the blue to the near
infrared. The upper two panels are U and B exposures respectively. In
both cases one can see the ``massive'' eastern (left) arm emphasizing
the asymmetry in the spiral structure of the dominant galaxy NGC~6907.
In the middle left (R-band) image the composite nature of the eastern
arm is beginning to become apparent as the elliptical form of NGC~6908
starts show itself. At progressively longer wavelengths (I and H-band
images in the lower two panels) the contrast between the (red)
companion and the (blue) spiral arm becomes even more enhanced and the
companion is obvious beyond question. The middle right panel is a
continuum-subtracted H$\alpha$ image of the system. For all intents
and purposes, the intruder, NGC~6908 has disappeared; but its effect
are still rather dramatically present in the obvious discontinuity in
the eastern arm delineated by the HII regions. At the position of
NGC~6908 there is an abrupt change in the pitch angle of the arm and
beyond that point the HII regions take up a distinctly linear
formation behind the intruder. Finally, it is noteworthy to point out
the finely delineated arc of HII regions that define the outermost
parts of the spiral structure of NGC~6907. They complete a full
360$\deg$ wrapping of the optically less prominent western arm. In
terms of bulk luminosity the eastern arm is dominant, but for total
angular extent the western arm is by far the most extensive. For an
expanded view of the H$\alpha$ image see Figure~3. A RGB map of the BR
and H-band images is shown in Figure~5.}
\end{figure}

\begin{figure}
\epsscale{1.0}
\plotone{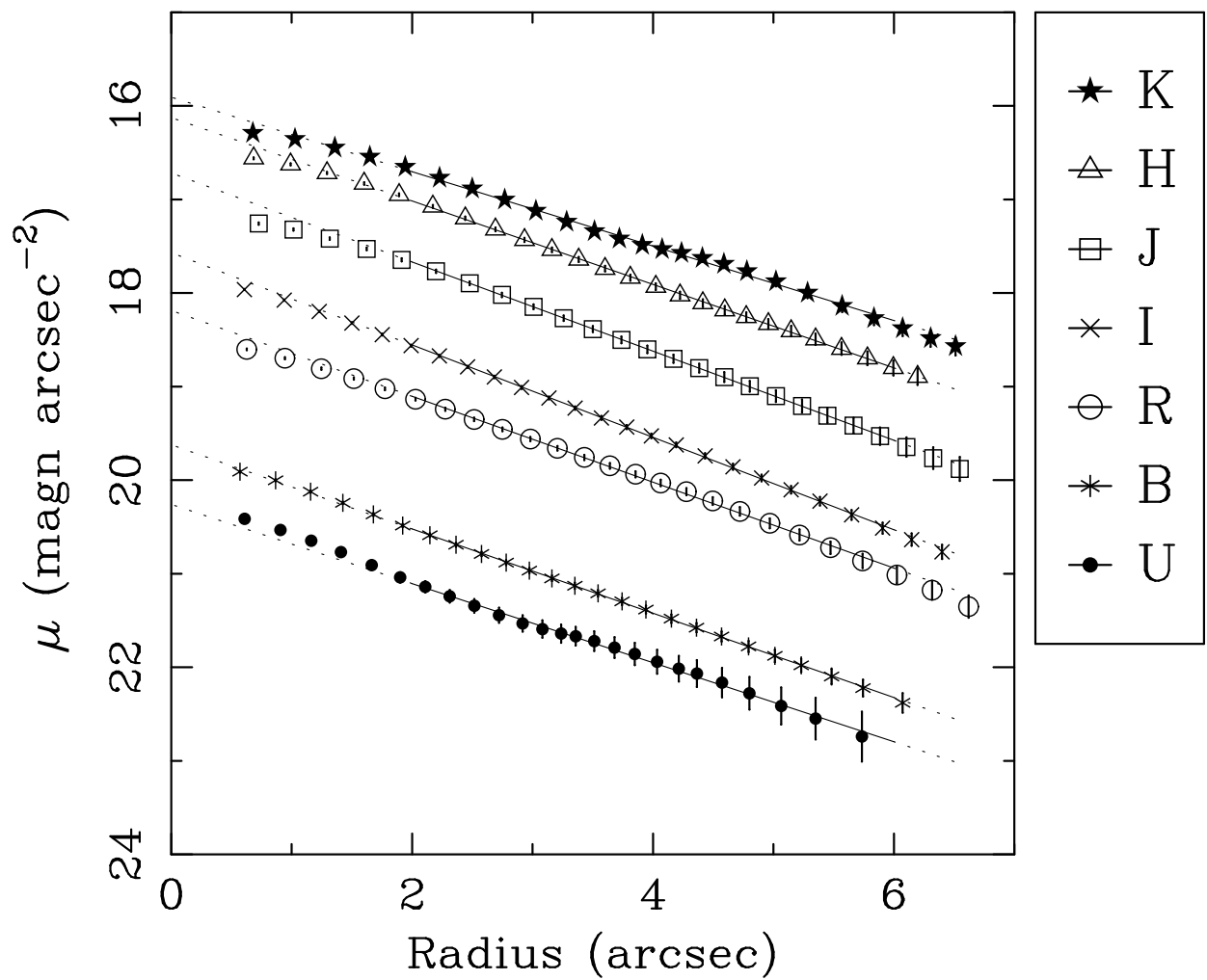}
\caption{This figure shows the multi-wavelength
radial light profiles of NGC~6908 derived from concentric elliptical
isophote fitting. As can be readily seen, the fall-off is exponential
from the near-ultraviolet (U-band: bottom profile) well out to 2.2
microns (K-band: top profile) confirming that this object is an S0
disk-like system, not an elliptical galaxy. Best-fitting straight
lines are shown (lines are shown solid within the ranges used for the
fit and dotted outside).}
\end{figure}
 
\begin{figure}
\plotone{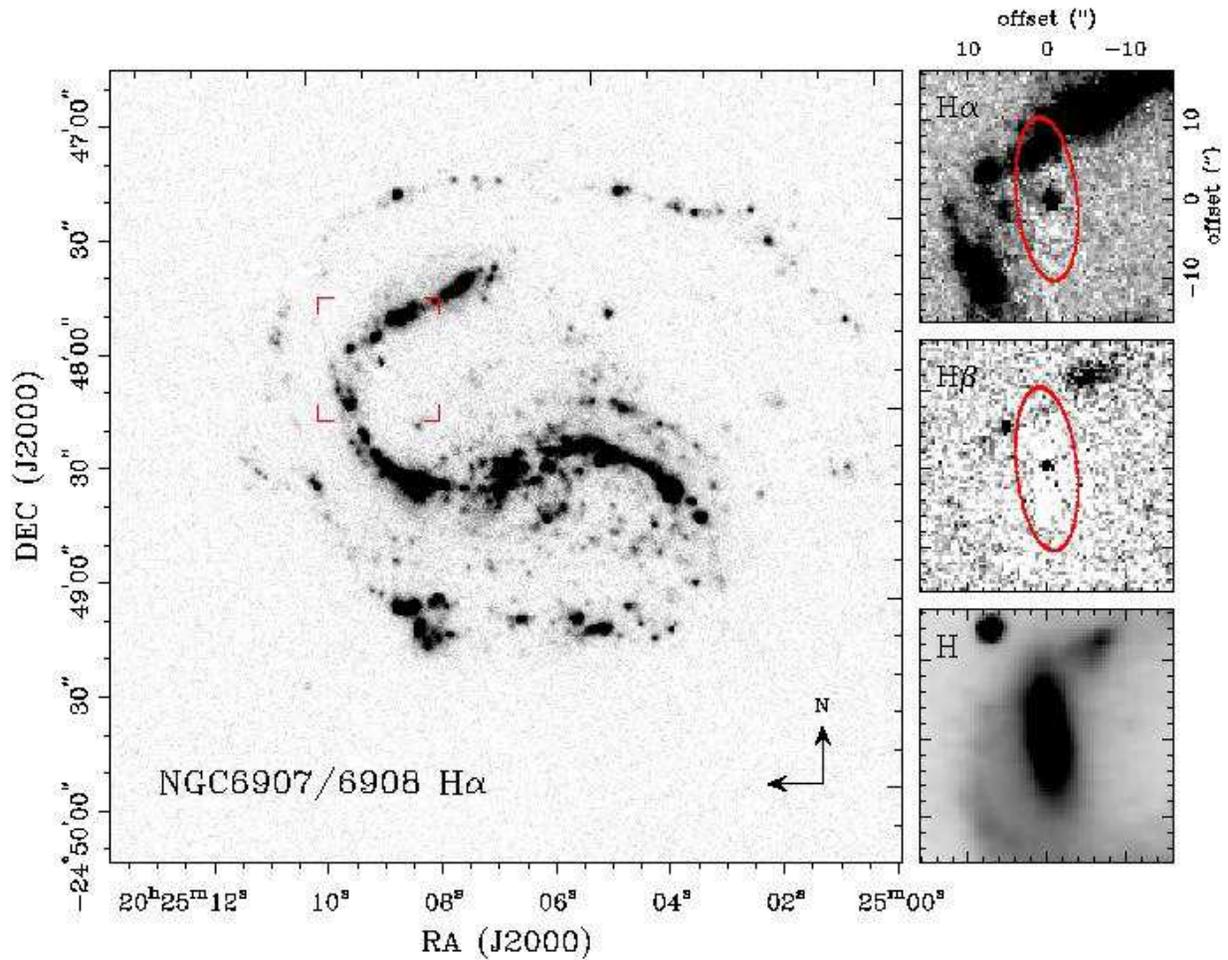}
\caption{H$\alpha$ image of NGC~6907/8. This
continuum-subtracted image was obtained from the Anglo-Australian
Telescope archive. The extent and the asymmetry of the spiral
structure is especially noteworthy. At the position of NGC~6907 the
image has been expanded (upper right panel) and two other wavelengths,
continuum-subtracted H$\beta$ (middle right panel) and the
near-infrared H-band image (lower right panel) are shown for
comparison. In the hydrogen-line images the overplotted ellipse shows
the continuum outline of the main body of NGC~6908.  Clearly the
intruder shows nuclear emission in H$\alpha$ and possibly in H$\beta$
too.}
\end{figure}
 
\begin{figure}
\plotone{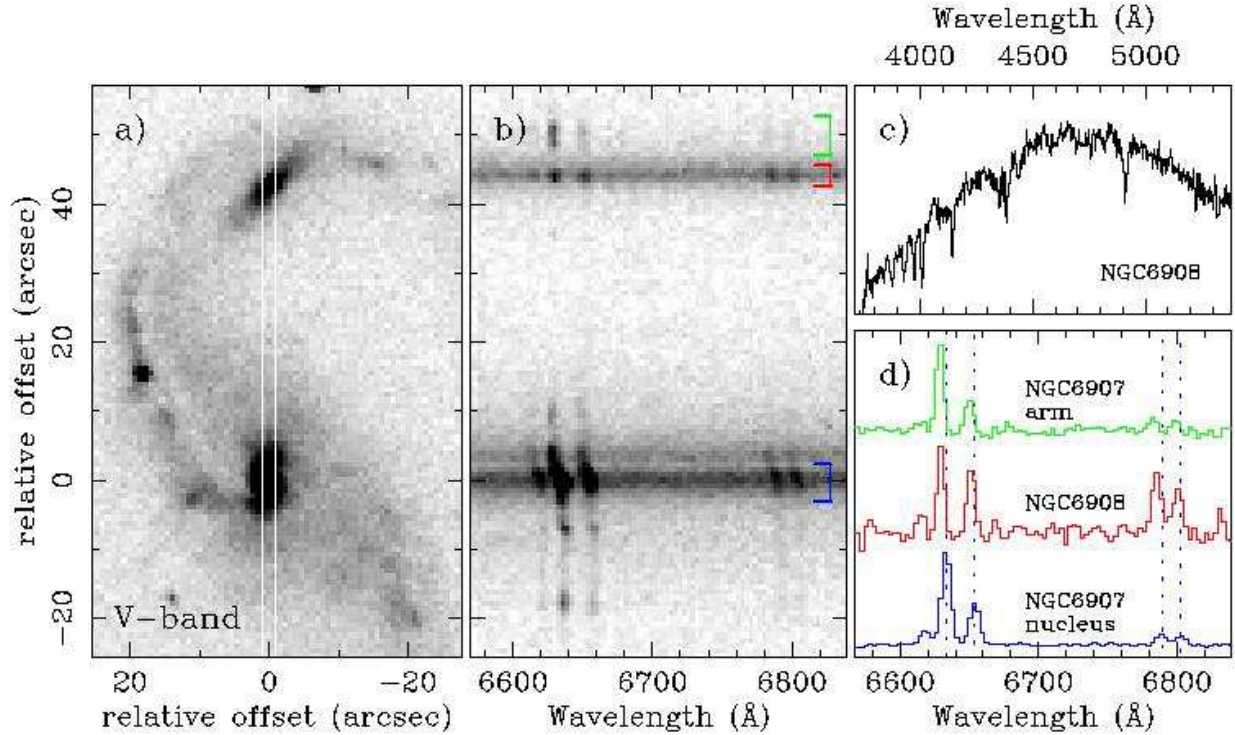}
\caption{Long-Slit Spectra of NGC~6907 and NGC~6908. The left panel
(a) shows the position of the 2-arcsec-wide slit used at the duPont
2.5m telescope superimposed on a V-band images of the two
galaxies. The central panel (b) shows the sky-subtracted region of the
spectrum (at the same spatial scale as panel a) used to determine the
radial velocities, expanded to show only the wavelength region around
H$\alpha$ and the two sulfur lines. Square brackets to the right show
the regions extracted for the radial velocity determinations,
including a nearby HII region just above NGC~6908. The upper right
panel (c) shows the (unflux-calibrated) hydrogen absorption-line
spectrum of NGC~6908 taken at the Magellan telescope; while the lower
right panel (d) shows the extracted emission-line spectra for the
three objects (obtained at the duPont telescope) where the dashed
vertical lines show the central wavelengths of H$\alpha$, [NII], and
the two sulfur lines for the nucleus of NGC~6907 projected up and
across the lines for the other two objects whose velocities are
visibly lower. Each of the emission-line spectra are scaled to
H$\alpha$ so as to show the increased relative strength of the [NII]
and [SII] lines in the nucleus of NGC~6908.}
\end{figure}
 
\begin{figure}
\plotone{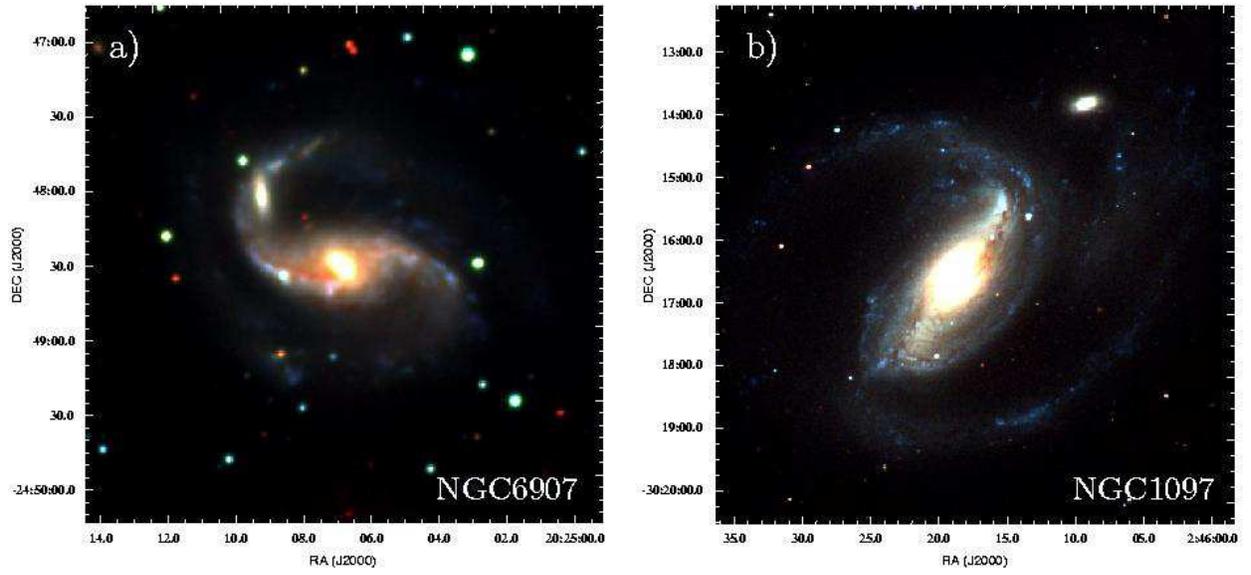}
\caption{RGB color composite image of NGC~6907/08 produced by combining H (red), 
R (green), and B-band (blue) images. Note the difference in color
between the North-West spiral arm of NGC~6907 and the S0 galaxy NGC~6908.}
\end{figure} 

\begin{figure}
\epsscale{1.0}
\plotone{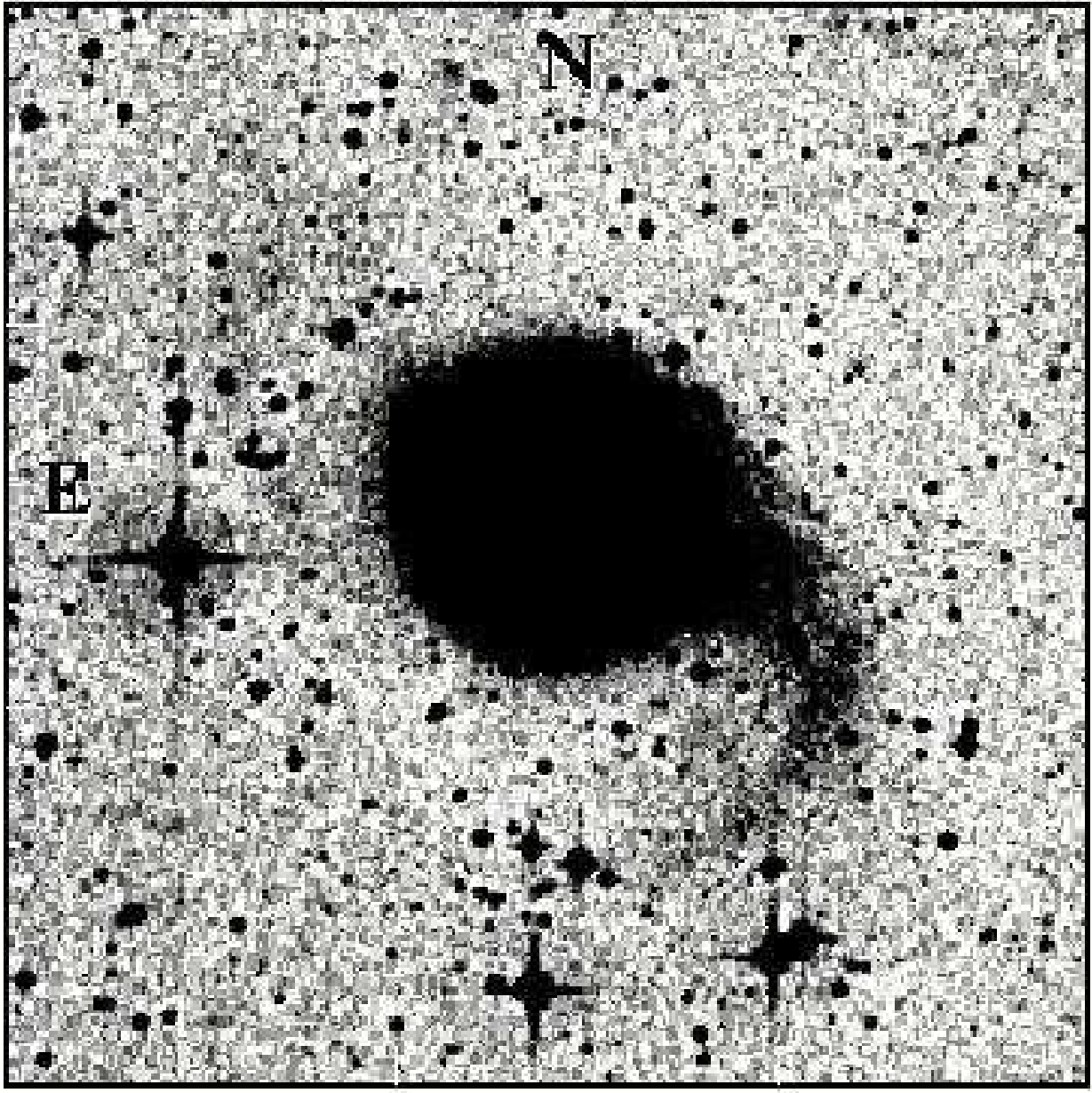}
\caption{High-contrast POSS image of the outer regions surrounding
NGC~6907/8. The one-armed spiral tail arcing from the north over to the
west is clearly visible in this image. A more diffuse region of tidal
debris is also visible to the north-east of the merging pair, arcing
over to the northern edge of the image. The full field of view is
10~arcmin.}
\end{figure}

\begin{figure}
\epsscale{1.0}
\plotone{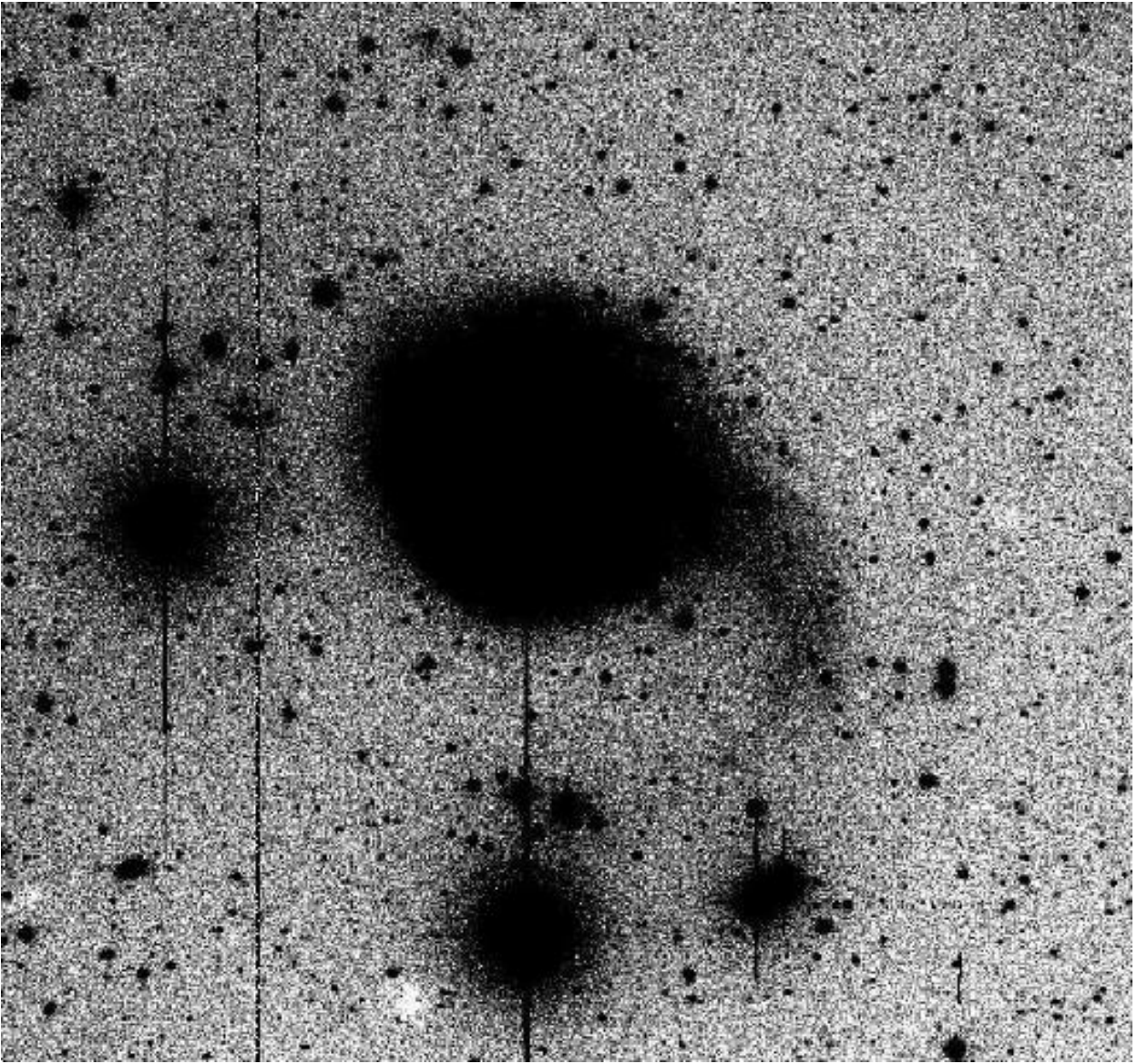}
\caption{R-band CCD image of NGC~6907/8 taken with the duPont 2.5m at
Las Campanas, Chile. The low-surface-brightness tidal feature to the
south-west of the main galaxy is confirmed in this independent image,
taken at a longer wavelength than the POSS (blue) image reproduced in
Figure 6.}
\end{figure}

\begin{figure}
\epsscale{0.65}
\plotone{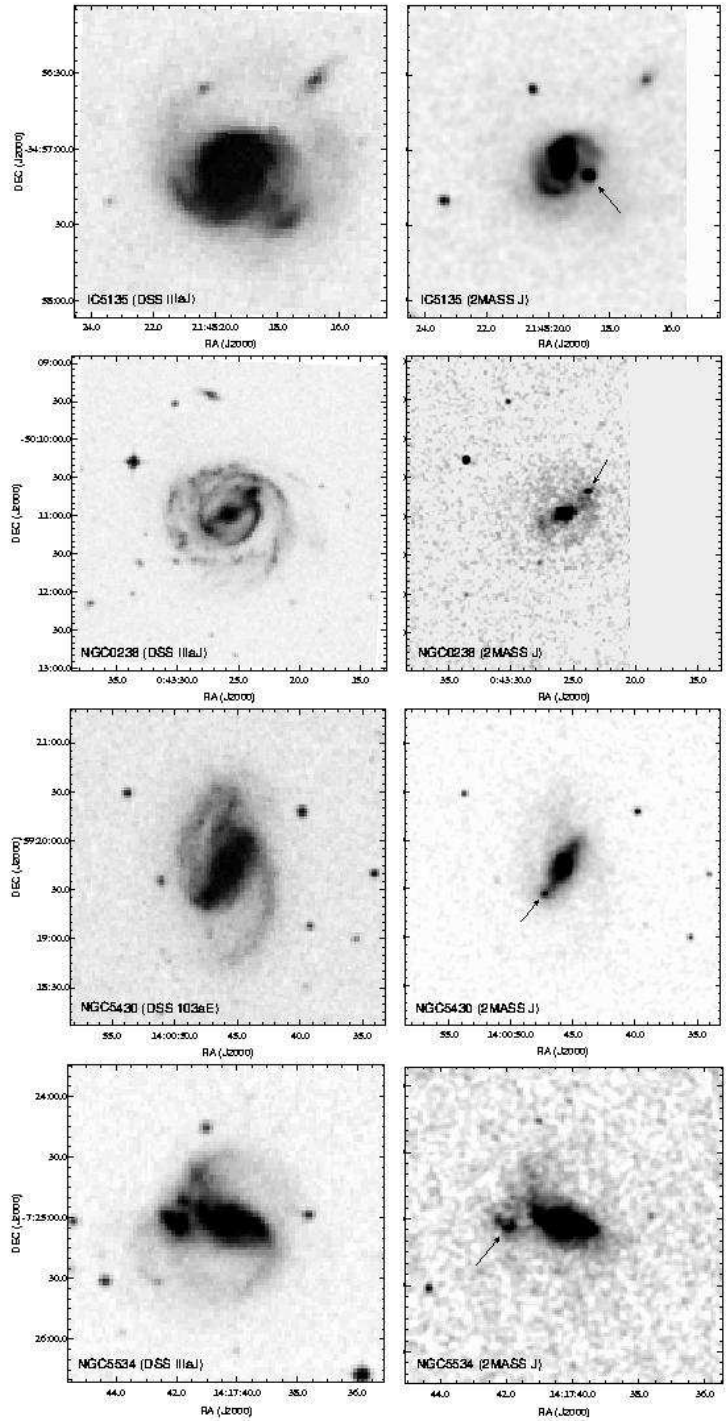}
\caption{Selected examples of embedded companions discovered using
long-wavelength archival images. The left-hand panels show optical images,
right-hand panels show the near-infrared discovery images.}
\end{figure}

\vfill\eject
\noindent

\end{document}